\documentclass[11pt]{article}
\usepackage{blois,epsfig}

\def\be{\begin{equation}}
\def\ee{\end{equation}}
\def\bea{\begin{eqnarray}}
\def\eea{\end{eqnarray}}

\usepackage{amsmath,amsfonts}
\newcommand{\ud}{\mathrm{d}}
\newcommand{\ve}{\varepsilon}
\long\def\symbolfootnote[#1]#2{\begingroup
\def\thefootnote{\fnsymbol{footnote}}\footnote[#1]{#2}\endgroup} 

\begin{document}
\vspace*{4cm}
\title{Scalar Field Cosmology -- Improving the Cosmological Evolutional Scenario}

\author{OREST HRYCYNA$^{a\hspace{0.5mm}}$\symbolfootnote[1]{Present address:
Theoretical Physics Division, National Centre for Nuclear Research, Ho{\.z}a 69,
00-681 Warszawa, Poland} and MAREK SZYD{\L}OWSKI$^{b,c}$}
\address{$^{a}$Department of Theoretical Physics,
The John Paul II Catholic University of Lublin, \\
Al. Rac{\l}awickie 14, 20-950 Lublin, Poland \\
$^{b}$Astronomical Observatory, Jagiellonian University, Orla 171, 30-244
Krak{\'o}w, Poland \\
$^{c}$Mark Kac Complex Systems Research Centre, Jagiellonian University, \\
Reymonta 4, 30-059 Krak{\'o}w, Poland}

\maketitle

\abstracts{
We study evolution of cosmological models filled with the scalar field and
barotropic matter. We consider the scalar field minimally and non-minimally
coupled to gravity. We demonstrated the growth of degree of complexity of
evolutional scenario through the description of matter content in terms of the
scalar field. In study of all evolutional paths for all initial conditions
methods of dynamical systems are used. Using linearized solutions we present
simple method of derivation corresponding form of the Hubble function of the
scale factor $H(a)$.}

The scalar fields play important role during the cosmic evolution. They 
are relevant in very early (quantum cosmology, inflation ) as well as 
late stages of evolution of the Universe (quintessence idea). We study the 
significance of scalar fields which are additionally non-minimally coupled to 
gravity in the phase of accelerated expansion of the Universe. For this 
aims we adopt the methods of dynamical systems which offers possibility 
of global investigation all evolutional paths for all admissible initial conditions.

We assume the Friedmann-Robertson-Walker (FRW) universe with an arbitrary curvature 
filled with the non-minimally coupled scalar field and barotropic fluid with the equation of
state coefficient $w_{m}$. The action integral is
\begin{equation}
S = \frac{1}{2\kappa^{2}}\int\ud^{4}x\sqrt{-g}R -
\frac{1}{2}\int\ud^{4}x\sqrt{-g}\Big(\ve\nabla^{\alpha}\phi\,\nabla_{\alpha}\phi
+\ve \xi R \phi^{2} + 2U(\phi)\Big) + S_{m}
\end{equation}
where $\kappa^{2}=8\pi G$, $\ve = \pm1$ corresponds to canonical and phantom
scalar fields, respectively, $R=6\big(\dot{H}+2H^{2}+k/a^{2}\big)$ is the Ricci
curvature scalar, and 
$$
U(\phi) = U_{0}\big(\phi^{2}-\phi_{0}^{2}\big)^{2}
$$
is the assumed form of the Higgs potential function. $S_{m}$ is the action for
the barotropic matter content.
From the Einstein equations we obtain the following forms of the energy conservation condition 
\begin{equation}
\frac{3}{\kappa^{2}}\left(H^{2}+\frac{k}{a^{2}}\right) =
\ve\frac{1}{2}\dot{\phi}^{2} + U(\phi) +\ve6\xi H
\phi\dot{\phi}+\ve3\xi\phi^{2}\left(H^{2}+\frac{k}{a^{2}}\right) + \rho_{m}\, ,
\label{eq:con1}
\end{equation}
and the acceleration equation
\begin{eqnarray}
\left(1-\ve\xi(1-6\xi)\kappa^{2}\phi^{2}\right)\dot{H}& = &  \frac{k}{a^{2}} -
\frac{\kappa^{2}}{2}\bigg[\ve(1-2\xi)\dot{\phi}^{2}+\ve8\xi H
\phi\dot{\phi}+2\xi\phi
U'(\phi)\nonumber \\ & & +
\ve2\xi(1+6\xi)\phi^{2}\frac{k}{a^{2}}+\ve24\xi^{2}\phi^{2}H^{2}+\rho_{m}(1+w_{m})\bigg]\,
.
\label{eq:acc1}
\end{eqnarray}
In what follows we introduce the energy phase space variables
$x\equiv \frac{\kappa \dot{\phi}}{\sqrt{6}H}$,
$y\equiv\frac{\sqrt{12\, U_{0}}}{\kappa}\frac{1}{H}$,
$z\equiv\frac{\kappa}{\sqrt{6}}\phi$,
in which the energy conservation condition (\ref{eq:con1}) and the acceleration equation
(\ref{eq:acc1}) are
\begin{equation}
\Omega_{m} = 1 -
\left[y^{2}(z^{2}-z_{0}^{2})^{2}+\ve(1-6\xi)x^{2}+\ve6\xi(x+z)^{2}+(1-\ve6\xi
z^{2})\Omega_{k}\right]
\end{equation}
\begin{eqnarray}
\left(1-\ve6\xi(1-6\xi)z^{2}\right) \frac{\dot{H}}{H^{2}} & = &
-\left(1-\ve6\xi(1+6\xi)z^{2}-\frac{3}{2}(1+w_{m})\left(1-\ve6\xi
z^{2}\right)\right)\Omega_{k} \nonumber \\ & &
\hspace{-10mm} -\ve\frac{3}{2}(1-w_{m})(1-6\xi)x^{2}-\ve3\xi(1-3w_{m})(x+z)^{2}
+ \ve12\xi(1-6\xi)z^{2} \nonumber \\ & & \hspace{-10mm}
-\frac{3}{2}(1+wm)\left(1-y^{2}(z^{2}-z_{0}^{2})^{2}\right) -12\xi
y^{2}z^{2}(z^{2}-z_{0}^{2})
\end{eqnarray}
where $\Omega_{m}=\frac{\kappa^{2}\rho_{m}}{3 H^{2}}$ and
$\Omega_{k}=-\frac{k}{a^{2}H^{2}}$.

\begin{figure}[t]
\begin{center}
\includegraphics[scale=0.6]{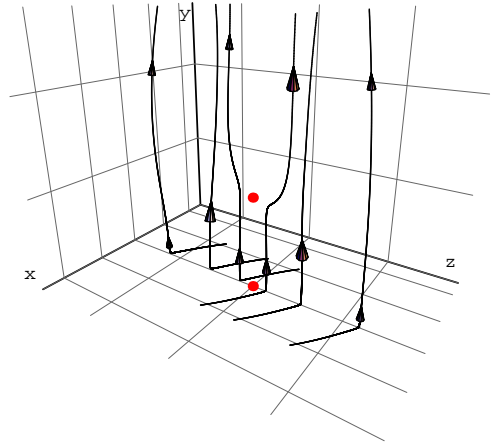}
\end{center}
\caption{The phase space portrait for the flat ($k=0$) FRW model filled with the
minimally coupled canonical scalar
field ($\ve=+1$, $\xi=0$) and the Higgs potential function
$U(\phi)=U_{0}\big(\phi^{2}-\phi_{0}^{2}\big)^{2}$ and barotropic dust matter
($w_{m}=0$). The critical points
correspond to the matter domination and transient acceleration epochs. All
trajectories asymptotically approach critical points located at $x^{*}=0$,
$y^{*}=\infty$ and $z^{*}= \pm z_{0}$
which corresponds to the Einstein universe with $H^{2} \to 0$.}
\label{fig:1}
\end{figure}

\begin{figure}[t]
\begin{center}
\includegraphics[scale=0.6]{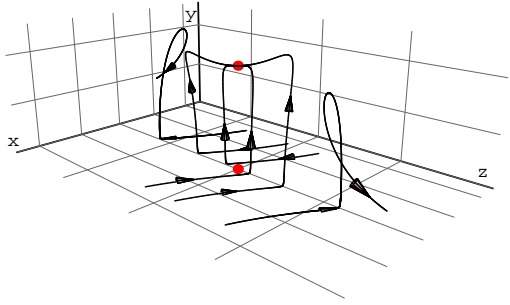}
\end{center}
\caption{The phase space portrait for the flat ($k=0$) FRW model filled with the minimally coupled phantom scalar field
($\ve=-1$, $\xi=0$) and the Higgs potential function
$U(\phi)=U_{0}\big(\phi^{2}-\phi_{0}^{2}\big)^{2}$ and barotropic dust matter
($w_{m}=0$). The red dots denote critical points of the system and correspond to
matter domination and accelerated expansion epochs. }
\label{fig:2}
\end{figure}

The dynamics of the model is completely described by the $4$-dimensional
dynamical system in variables $x$, $y$, $z$ and $\Omega_{k}$
\begin{eqnarray}
\label{eq:system1}
x' & = & -2\left(1-\ve6\xi(1-6\xi)z^{2}\right)\left(x+3\xi z+\ve
y^{2}z(z^{2}-z_{0}^{2})\right) \nonumber \\ & & - (x+6\xi
z)\bigg[1+\ve6\xi(1-6\xi)z^{2}-\ve\frac{3}{2}(1-w_{m})(1-6\xi)x^{2}-\ve3\xi(1-3w_{m})(x+z)^{2}
\nonumber \\ & & \hspace{20mm} -\frac{3}{2}(1+w_{m})(1-y^{2}(z^{2}-z_{0}^{2})^{2}) - 12\xi
y^{2}z^{2}(z^{2}-z_{0}^{2})\bigg] \nonumber \\ & & + \frac{1}{2}(x+6\xi
z)(1-3w_{m})(1-\ve6\xi z^{2})\Omega_{k} - x(1-\ve6\xi(1-6\xi)z^{2})\Omega_{k} \,
\label{eq:system2}
\end{eqnarray}
\begin{eqnarray}
y' & = & y\bigg[\big(1-\ve6\xi(1+6\xi)z^{2}-\frac{3}{2}(1+w_{m})(1-\ve6\xi
z^{2})\big)\Omega_{k}
+\ve\frac{3}{2}(1-w_{m})(1-6\xi)x^{2} \nonumber \\ & & \hspace{5mm}
+\ve3\xi(1-3w_{m})(x+z)^{2} -
\ve12\xi(1-6\xi)z^{2}+\frac{3}{2}(1+w_{m})\big(1-y^{2}(z^{2}-z_{0}^{2})^{2}\big)
\nonumber \\ & & \hspace{5mm}
+12\xi y^{2}z^{2}(z^{2}-z_{0}^{2})\bigg]\,\\
\label{eq:system3}
z' & = & x\left(1-\ve6\xi(1-6\xi)z^{2}\right)\, ,\\
\label{eq:system4}
\Omega_{k}' & = &
2\Omega_{k}\bigg[\big(1-\ve6\xi(1+6\xi)z^{2}-\frac{3}{2}(1+w_{m})(1-\ve6\xi
z^{2})\big)\Omega_{k}
+\ve\frac{3}{2}(1-w_{m})(1-6\xi)x^{2} \nonumber \\ & & \hspace{5mm}
+\ve3\xi(1-3w_{m})(x+z)^{2} - 1- 
\ve6\xi(1-6\xi)z^{2}+\frac{3}{2}(1+w_{m})\big(1-y^{2}(z^{2}-z_{0}^{2})^{2}\big)
\nonumber \\ & & \hspace{5mm}
+12\xi y^{2}z^{2}(z^{2}-z_{0}^{2})\bigg]
\end{eqnarray}
where the differentiation is with respect to time $\tau$ defined as
$\frac{\ud}{\ud\tau}=\left(1-\ve6\xi(1-6\xi)z^{2}\right)\frac{\ud}{\ud\ln{a}}$.
On figures \ref{fig:1} and \ref{fig:2} we present the phase space portraits
for the special cases of the minimally ($\xi=0$) coupled scalar fields, both,
canonical ($\ve=+1$) and phantom ($\ve=-1$). The dynamics of non-minimally
coupled scalar fields is different because of the coupling between the Ricci
curvature scalar and the scalar field \cite{Hrycyna:2009zj,Hrycyna:2010yv}.

In what follows we concentrate on the single critical point which corresponds to
the accelerated expansion of the universe, namely the critical point located at
($x^{*}=0$, $y^{*}=1/z_{0}^{2}$, $z^{*}=0$, $\Omega_{k}^{*}=0$). Using the
linearised solution in the vicinity of this point we will show that the solution
form of the Hubble function corresponds to the $\Lambda$CDM model independent of the
values of the $\xi$ and $z_{0}$ parameters.

Using the energy phase space variable
$y\equiv\frac{\sqrt{12\,U_{0}}}{\kappa}\frac{1}{H}$ we can write that
\begin{equation}
\frac{H^{2}}{H_{0}^{2}} = \frac{y(a_{0})^{2}}{y(a)^{2}}
\end{equation}
where $H_{0}$ and $y(a_{0})$ are the present time values. Only the
linearised solution of the variable $y(\tau)$ is needed in derivation of the
Hubble function. It is in the following form
\begin{equation}
y(\tau) = y^{*} + \frac{1}{2 z_{0}^{2}}\big(2z_{0}^{2}\Delta y +
\Delta\Omega_{k}\big)\,\,e^{-3(1+w_{m})\tau} -
\frac{1}{2z_{0}^{2}}\Delta\Omega_{k}\,\,e^{-2\tau}\,,
\end{equation}
where $\Delta y = y^{*} - y^{(i)}$ and
$\Delta\Omega_{k}=\Omega_{k}^{*}-\Omega_{k}^{(i)}$ are the initial conditions.
Up to linear terms we have $z(\tau)^{2}\approx0$, then we obtain that the time
$\tau$ transforms in to the scale factor as follows
$\tau\approx\ln{(\frac{a}{a^{(i)}})}$ where $a^{(i)}$ is the initial value of
the scale factor. Finally
\begin{eqnarray}
\frac{y(a_{0})^{2}}{y(a)^{2}} & \approx & 1 + (2 z_{0}^{2}\Delta y +
\Delta\Omega_{k})\left(\frac{a_{0}}{a^{(i)}}\right)^{-3(1+w_{m})} -
\Delta\Omega_{k}\left(\frac{a_{0}}{a^{(i)}}\right)^{-2} \nonumber \\ & &- (2 z_{0}^{2}\Delta y +
\Delta\Omega_{k})\left(\frac{a}{a^{(i)}}\right)^{-3(1+w_{m})} +
\Delta\Omega_{k}\left(\frac{a}{a^{(i)}}\right)^{-2}.
\end{eqnarray}
From the energy conservation condition up to linear terms we have
$\Omega_{m,i}\approx-(2z_{0}^{2}\Delta y + \Delta\Omega_{k})$
hence we obtain
\begin{equation}
\frac{H^{2}}{H_{0}^{2}} = 1 - \Omega_{m,0} - \Omega_{k,0} +
\Omega_{m,0}\left(\frac{a}{a_{0}}\right)^{-3(1+w_{m})} +
\Omega_{k,0}\left(\frac{a}{a_{0}}\right)^{-2}
\end{equation}
which for the dust matter $w_{m}=0$ represents the $\Lambda$CDM model with the
curvature.
\begin{figure}
\begin{center}
\includegraphics[scale=0.4]{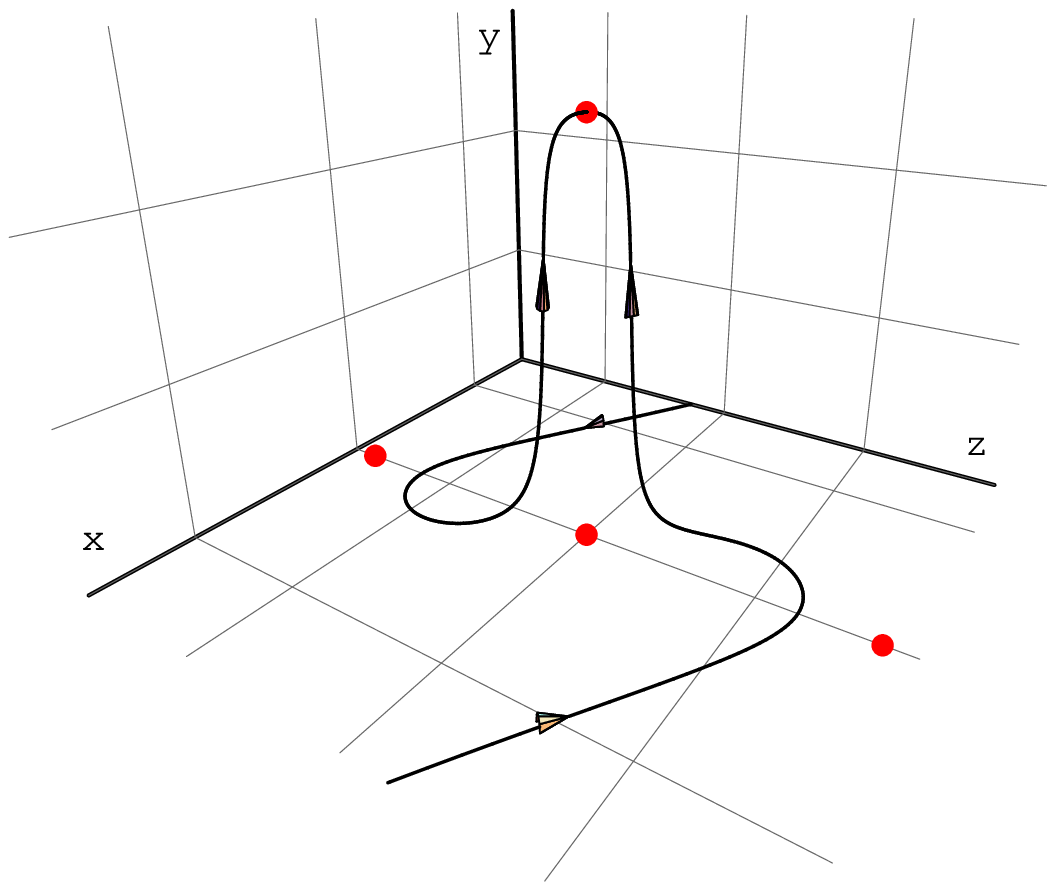}
\includegraphics[scale=0.6]{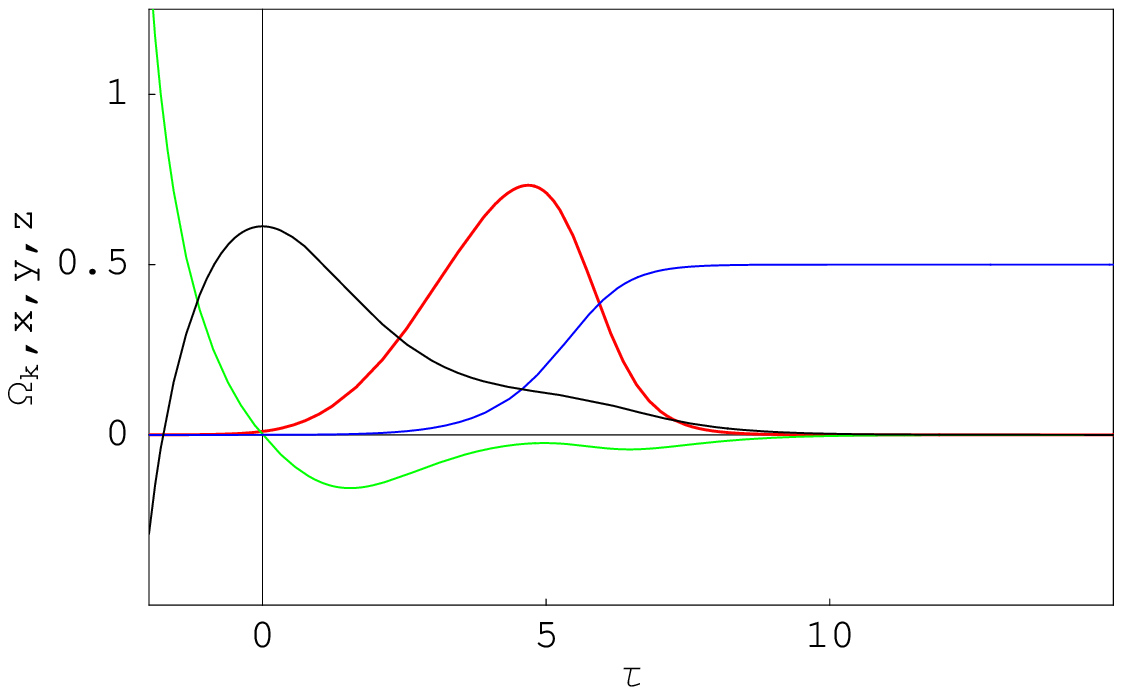}
\end{center}
\caption{The evolution described by the system of equations
(\ref{eq:system1}--\ref{eq:system4}) with $\xi=1/4$ and $z_{0}^{2}=2$. On the
left panel we present the projection of $4$-dimensional phase space on the
$3$-dimensional subspace ($x$, $y$, $z$) and on the right panel we present time
evolution of the phase space variables: $\Omega_{k}(\tau)$ -- red, $x(\tau)$ --
green, $y(\tau)$ -- blue, $z(\tau)$ -- black. The sample trajectories
interpolate between three major epochs in the history of the universe: the
radiation dominated universe, the barotropic matter and finally the
quintessence domination epoch.}
\label{fig:3}
\end{figure}

The dynamics of the scalar field with non-minimal coupling to gravity was studied 
in details for the Higgs potential of the scalar fields in the framework of 
dynamical systems methods. We construct 3-dimensional phase portraits for flat 
and non-flat cosmological models. It is additionally assumed the presence of barotropic matter 
in the model which gives rise to a new evolutional scenario and enlarges a degree 
of complexity of cosmic evolution. The structure of phase space was investigated. 
From the linearization of the system around the critical points we obtain the 
crucial formula for $H^2$ which demonstrated how the standard cosmological model 
emerges from the model under consideration.

\section*{References}

\bibliography{../../darkenergy,../../quintessence}

\end{document}